\documentclass[aps,prb,showpacs,reprint]{revtex4-1}

\usepackage{graphicx}
\usepackage{amsmath}
\usepackage{bbm}

\begin{document}
\title
{Spin filter effect at room temperature in GaN/GaMnN ferromagnetic resonant tunnelling diode.}
\author{P. W\'ojcik}
\email[Electronic address: ]{Pawel.Wojcik@fis.agh.edu.pl}
\author{J. Adamowski}
\author{M. Wo{\l}oszyn}
\author{B.J. Spisak}
\affiliation{University of Science and Technology,
Faculty of Physics and Applied Computer Science,
Krak\'ow, Poland}

\begin {abstract}
We have investigated the spin current polarization without the external magnetic field
in the resonant tunneling diode with the emitter and quantum well layers made from the
ferromagnetic GaMnN. For this purpose we have applied the self-consistent Wigner-Poisson
method and studied the spin-polarizing effect of the parallel and antiparallel alignment of
the magnetization in the ferromagnetic layers. The results of our calculations show that the
antiparallel magnetization is much more advantageous for the spin filter operation and leads
to the full spin current polarization at low temperatures and 35 \% spin polarization of the
current at room temperature.
\end{abstract}

\maketitle
The progress in homo- and heteroepitaxy of dilute magnetic 
semiconductors\cite{Koo2010,Agarwal2006,Edmonds2002,Yu2002,Chiba2003,Ku2003,Eid2005,Jungwirth2005}
(DMS's) during the past decade allows to fabricate spintronic nanodevices, in which the spin 
polarization of the current can be controlled by the magnetic or electric field. 
The spin filter effect in a resonant tunneling diode (RTD) with paramagnetic quantum well 
embedded in II-VI DMS (ZnMnSe) was studied  theoretically\cite{Wojcik2012,Wojcik2012JAP,Wojcik2012SST} 
and experimentally demonstrated by Slobodskyy~et~al.\cite{Slobodskyy2003}
For the paramagnetic RTD, in the presence of the external magnetic field, 
the exchange interaction between the conduction band electrons and the Mn$^{2+}$ ions leads 
the giant Zeeman splitting\cite{Furdyna1988} of the quasi--bound state in the paramagnetic 
quantum-well. This splitting causes that the resonance conditions for the spin up and spin 
down electrons are satisfied for different bias voltages leading to the separation of the 
spin current components and consequently to the spin polarization of the current. 
The spin filter effect in the paramagnetic RTD is limited to very low temperature 
and requires the high external magnetic field.\cite{Slobodskyy2003,Furdyna1988} 
These restrictions cause that more interest is directed towards the application of the 
ferromagnetic III-V semiconductors, especially those with high Curie temperature, 
e.g.~GaMnAs\cite{Jungwirth2005,Wang2005,Panguluri2004,Csontos2005} or GaMnN.\cite{Dietl2000,Liu2005}
Ohno~et~al.\cite{Ohno1998} experimentally studied the ferromagnetic RTD based on GaMnAs in which the spin 
splitting was observed without external magnetic field but it still requires low temperature. 
Hovewer, the recent experiments reported that GaMnN can exhibit
the ferromagnetic properties above room temperature\cite{Sasaki2002,Ando2003,Pearton2007} 
at which the exchange splitting of the conduction band is about few tens of meV\cite{Ertler2007} 
and remains in  the limit of thin layer of a few nanometer width.\cite{Priour2005} 
Although the ferromagnetism in GaMnN is still unresolved theoretical problem, 
the spin filter effect in the RTD's based on GaMnN is a subject of research carried out by many groups.
Recently, Li~et~al.\cite{Li2006} have theoretically investigated the ferromagnetic RTD consisted
of the InGaN quantum well between two GaMnN barriers. In Ref.~\onlinecite{Li2006} the spin polarization
of the current without magnetic field has been predicted at low temperature but at room temperature 
it has been reduced to only 8 \%. 
Another way to obtain the spin polarization of the current at room temperature
was proposed by Qui~el~at.\cite{Qui2008} who stated that the $\delta$ doping of the GaN quantum 
well in the RTD with ferromagnetic (GaMnN) contacts enhances the spin polarization of the current 
by two times. 
The double enhancement of the spin polarization at room temperature  was also 
reported by Wang~et~al.\cite{Wang2009} who investigated the influence of the charge polarization 
at the interface AlGaN/GaN  in RTD with ferromagnetic  
contacts embedded in GaMnN.

In the present paper, we propose the RTD structure with the ferromagnetic emitter and 
quantum well regions made from GaMnN. Based on self-consistent Wigner-Poisson calculations we
predict the full spin polarization of the current (i.e. $P=\pm 1$) at low temperature for antiparallel 
magnetization of the magnetic layers.
The spin polarization reduces to $P=\pm 0.35$ at room temperature. To the best of our knowledge, 
this is the highest value  of the spin polarization predicted at room temperature in magnetic RTD. 
In particular it is about four times higher than that reported by Li~et~al.~\cite{Li2006} 
We also showe that the proposed ferromagnetic RTD structure with the antiparallel alignment 
of magnetization can lead to the fairly large spin polarization of the current at room temperature.	
\begin{figure}[ht]
\begin{center}
\includegraphics[scale=0.5]{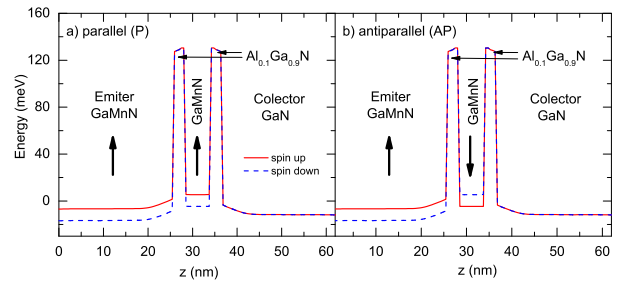}
\caption{Self-consistent potential energy profile for spin up and spin down electrons calculated
for a) parallel (P) and b) antiparallel (AP) alignment of the magnetization of the emitter and
quantum well layers.}
\label{fig1}
\end{center}
\end{figure}

We investigate the ferromagnetic RTD based on GaN/Al$_{1-x}$Ga$_x$N/GaMnN with 
the emitter and quantum well layer made from GaMnN (Fig.~\ref{fig1}). The parallel (P) and 
antiparallel (AP) alignments of the  magnetization of the emitter and quantum well
is considered. The conduction band profiles at zero bias for spin up and spin down electrons are 
presented in Fig.~\ref{fig1}.
The calculations have been performed with the following nanostructure parameters:
the thickness of the GaMnN quantum well layer is 6 nm, the barriers are assumed to be symmetric 
with thickness 3 nm and $x=0.1$ yields the barrier height 130~meV. ~\cite{Ambacher2002}
We assume the conduction-band electrons of GaN, 
i.e., m/m$_0$ = 0.228 and take on the relative  electric permittivity $\epsilon$ = 8.6. 
In order to describe the pure spin filter effect in the proposed ferromagnetic nanostructure we 
neglect the charge polarization occurring at the interface AlGaN/GaN.\cite{Ambacher2002}

Our numerical calculations are based on the Wigner-Poisson approach, according to which
the conduction band electrons are described by the spin dependent Wigner distribution
function (WDF).
Assuming the translational invariance in the $x-y$ plane and neglecting the spin 
scattering (many transport experiments showed that the spin scattering lenght 
is compared to the size of the RTD\cite{Ohno1998,Ohya2005})
the time independent quantum transport 
equations can be reduced to the following one-dimensional form:\cite{Spisak2009}
\begin{equation}
\frac{\hbar k}{m}\frac{\partial \rho^\mathcal{W}_{\sigma}(z,k)}{\partial z}
 = \frac{i}{2\pi \hbar}\int\limits_{-\infty}^{+\infty}{\mathrm{d}k^{\prime}}
 \mathcal{U}_{\sigma}(z,k-k^{\prime}) \rho^\mathcal{W}_{\sigma}(z, k^{\prime}),
\label{eq:WEq}
\end{equation}
where $\rho^\mathcal{W}_{\sigma}(z,k)$ is the spin-dependent WDF, $k$ is the
$z$-component of the wave vector and $\sigma=(\uparrow, \downarrow)$ is the spin index.\\
The non-local potential $\mathcal{U}_{\sigma}(z,k-k^{\prime})$ in Eq.~(\ref{eq:WEq}) is 
given by the formula
\begin{equation}
 \mathcal{U}_{\sigma}(z,k-k^{\prime})  = 
\int\limits_{-\infty}^{+\infty} {\mathrm{d}z^{\prime}}~\big[U_{\sigma}(z+\frac{z^{\prime}}{2})
 -  U_{\sigma}(z-\frac{z^{\prime}}{2})\big] e^{-i(k-k^{\prime})z^{\prime}},
\end{equation}
where $U_{\sigma}(z)$ is the spin-dependent potential energy profile, which
can be expressed as the sum of the three terms
$U_{\sigma}(z)=U_{\sigma}^0(z)+U_{el}(z)+U^{ex}_{\sigma}(z)$, 
where consecutive terms denote the spin-dependent conduction-band potential energy, 
the electrostatic potential energy calculated by solving the Poisson equation 
and the exchange energy.\\
Equations (\ref{eq:WEq}) and the Poisson equation form the system of non-linear integro-differential equations
that is solved by the self-consistent procedure.\cite{Wojcik2012}
After reaching convergence the spin dependent current density is calculated using the formula
\begin{equation}
j_{\sigma}=\frac{e}{2\pi L}\int\limits_0^L dz\int\limits_{-\infty}^{+\infty}{\mathrm{d}k}
\frac{\hbar k}{m}\rho_{\sigma}^\mathcal{W}(z,k), 
\end{equation}
where $L$ is the length of the nanodevice. 
\begin{figure}[h]
\begin{center}
\includegraphics[scale=0.6]{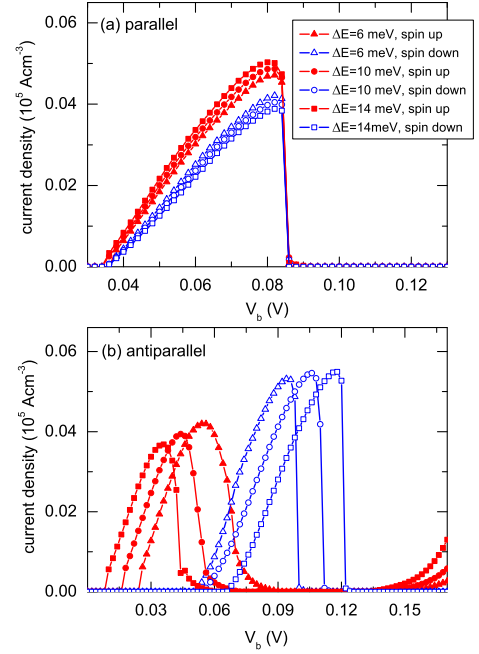}
\caption{Current-voltage characteristics for spin up (red) and spin down (blue) current components 
calculated for different values of the splitting energy $\Delta E$ and a) parallel, b) antiparallel 
alignment of the magnetization of the emitter and the quantum well layers.}
\label{fig2}
\end{center}
\end{figure}

The ferromagnetic properties of GaMnN
causes that the spin-degenerate quasi-bound state energy level in the quantum well 
splits into two levels for spin up and spin down electrons.
Similarly, the conduction band in the ferromagnetic emitter layer 
is splited into two subbands for different spins.
The spin splitting of the conduction bands in the ferromagnetic layers 
causes that the resonance transport conditions are different 
for the electrons with different spins.
In the present calculations the spin splitting energy $\Delta E$ of
of the conduction band  is treated as an parameter of calculations that
varies from 2~meV to 15~meV (the reliable values reported in experiments~\cite{Dietl2000,Liu2005}). 
Fig.~\ref{fig2} shows the spin-dependent current-voltage characteristics calculated at temperature $T=4.2$~K
for (a) parallel and (b) antiparallel alignment of the magnetization of
the emitter and the quantum well layers.
\begin{figure}[ht]
\begin{center}
\includegraphics[scale=0.6]{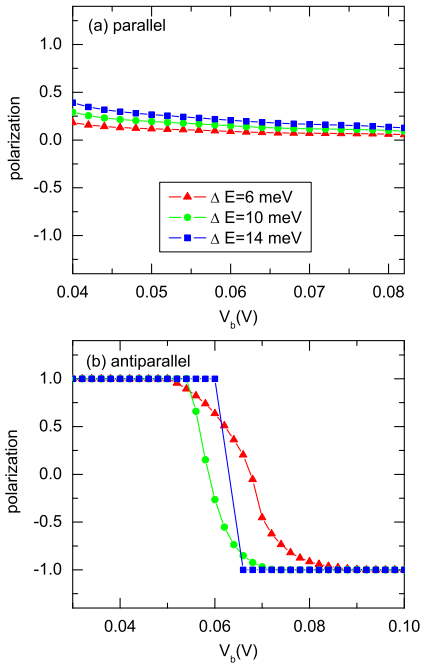}
\caption{Spin polarization of current $P$ as a function of bias $V_b$
for different value of the splitting energy $\Delta E$ and a) parallel and b) antiparallel 
alignment of the magnetization of the emitter and the quantum well layers.}
\label{fig3}
\end{center}
\end{figure}
We see that for the parallel magnetization of the layers the resonant current peaks
for spin up and spin down current component occur at almost the same bias $V_b$. 
If the splitting energy $\Delta E$ increases, the
resonant current peak increases for spin up and decreases for spin down current component. 
On the other hand for the antiparallel magnetization of the magnetic layers the resonant current peaks for both 
the spin components behave in a different manner. Namely, the increasing $\Delta E$ causes
the separation of the resonant current peaks: the resonant peak  for the spin up current component shifts towards 
the lower bias while the resonant peak  for the spin down current component shifts towards the higher bias.
The separation of the spin current components leads to the spin 
polarization of the current defined as 
$P=(j_{\uparrow}-j_{\downarrow})/(j_{\uparrow}+j_{\downarrow})$. 
In Fig.~\ref{fig3} we present the spin polarization of the current as a function of the bias calculated for
(a) parallel and (b) antiparallel alignment of the magnetization of the ferromagnetic layers.
We see that for the parallel magnetization the spin polarization of the current is positive at the
low bias and decreases with increasing the bias. 
On the other hand for the antiparallel magnetization of the ferromagnetic layers
the spin polarization of the current varies from $P=+1$
for the low bias to $P=-1$ for the high bias. 
This dependence is observed for all values of the splitting energy 
$\Delta E$, however, for the higher $\Delta E$ the transition between 
both the fully polarized states occurs in a narrower bias range.

\begin{figure}[ht]
\begin{center}
 \includegraphics[scale=0.32]{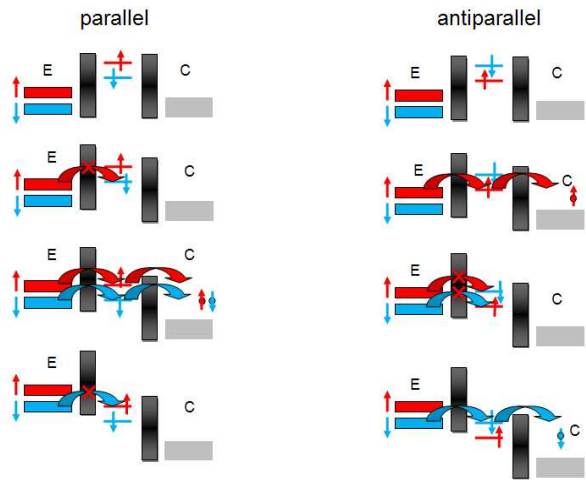}
\caption{Schematic illustration of spin dependent resonance tunneling of electrons in ferromagnetic RTD.
Only the antiparallel magnetization of the ferromagnetic layers leads to the full spin polarization of the current.
E and C denotes emitter and colector region. The bias increases from top to bottom panels.}
\label{fig4}
\end{center}
\end{figure}

\begin{figure}[ht]
\begin{center}
\includegraphics[scale=0.6]{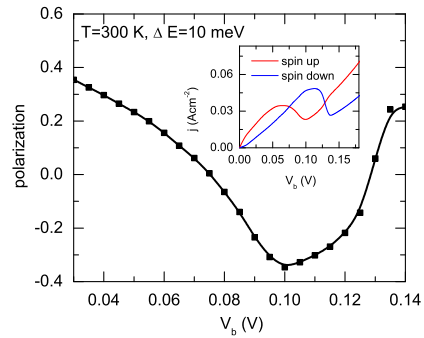}
\caption{Spin polarization of current as a function of  bias 
for antiparallel  alignment of the magnetization between the emitter and the quantum well
at room temperature $T=300$~K. Inset: current-voltage characteristics for spin up and spin down current components 
at room temperature.}
\label{fig5}
\end{center}
\end{figure}

In order to explain strongly polarizing effect of the ferromagnetic RTD with the antiparallel 
magnetization of the ferromagnetic layers we present the simple model of the spin dependent electron transport
through the RTD for both alignments of the magnetization (Fig.~\ref{fig4}).
Fig.~\ref{fig4} shows that only the antiparallel alignment of the 
magnetization can lead to the full spin polarization of the current.
Let us note that at room temparature the transport window in the magnetic 
emitter broadens in the nearest of the Fermi energy. This thermal effect causes the
broadenning of the resonant current peak for spin up and spin down current components
(inset in Fig.~\ref{fig5}). Our calculations show that even for small but experimental realiable value 
of the splitting energy $\Delta E=10$~meV the spin polarization at room temperature is still quite 
large and achieves $P=0.35$ (Fig.~\ref{fig5}). This value
is four times larger than that reported in Ref.~\onlinecite{Li2006}. 
Moreover we expect that for the larger splitting energy $\Delta E$, 
the antiparallel magnetization of the magnetic layers can
lead to the full spin polarisation of the current at room temperature.

In conclusion, we have shown that the antiparallel alignment of the 
magnetization in the ferromagnetic resonant tunneling structure with the ferromagnetic 
emitter and quantum well can be used to obtain the full spin polarization of the current 
at room temperature. Our theoretical calculations predicts that the spin polarization of the current 
in the ferromagnetic RTD based on GaMnN achieves $P=0.35$ at room temperature for experimentally reported 
splitting energy $\Delta E=10$~meV in GaMnN. We also argue that proposed nanostructure can allow to increase the 
polarization up to $|P|=1$ at room temperature for sufficiently large splitting energy $\Delta E$.
The achievement of the full spin current polarisation at room temperature by increasing of $\Delta E$ is 
a challenge for the future  spintronic technology which allows to construct 
the effective spin filter working at room temperature.

This paper has been supported by the Polish Ministry
of Science and Higher Education and its grants for Scientific
Research.

\end{document}